\documentclass[11pt]{article}
\usepackage[T1]{fontenc}
\usepackage{lmodern}
\usepackage{graphicx}
\usepackage{amsmath,amssymb,amsfonts}
\usepackage{booktabs}
\usepackage[hidelinks]{hyperref}
\usepackage[margin=1in]{geometry}
\begin{document}

\title{Trace-distance characterization of recurrent pairwise synchronization in cavity-coupled superconducting qubits}
\author{Tian-tian Huan\\
School of Mathematics and Computer Science, Chifeng University, Chifeng 024000, China\\
\texttt{huantt5956@163.com}}
\date{}

\maketitle

\begin{abstract}
We theoretically investigate recurrent pairwise synchronization in three superconducting qubits coupled to a common single-mode microwave cavity. The qubits interact indirectly through the cavity field and are driven by a coherent microwave field. We obtain the coherent dynamics in the dressed-state basis and quantify pairwise synchronization by the trace distance between reduced single-qubit density matrices. The trace-distance dynamics exhibit oscillatory wave-packet structures, with low-trace-distance regions identifying finite synchronization windows. An envelope-threshold procedure is introduced to extract the representative duration of these windows. By varying the middle-qubit coupling strength and the relative coupling ratio of the outer qubits, we construct a parameter-dependent map of the synchronization-window duration. The duration decreases rapidly with increasing relative outer-qubit coupling and becomes weakly dependent on the middle-qubit coupling in the stronger-coupling regime. These results provide a reduced-state-based characterization of coherent collective dynamics in multiqubit circuit-QED systems.
\end{abstract}


\section{Introduction}\label{sec:introduction}

Quantum synchronization has become an important topic at the intersection of nonlinear dynamics and quantum information science. In classical systems, synchronization typically describes the adjustment of the rhythms of coupled oscillators through mutual interaction or external driving. However, synchronization cannot be characterized solely by classical phase locking in quantum systems. To describe synchronization in the quantum regime, researchers have proposed various theoretical measures, including mutual synchronization, phase-space measures, complete synchronization measures, and state-distance-based approaches~\cite{Giorgi2012,Mari2013,Lee2013,Walter2014,Galve2017}. Furthermore, engineered protocols such as shortcuts to synchronization have shown that the time scale of synchronization can be actively manipulated in both classical and quantum systems~\cite{Impens2023}. Nevertheless, a universal definition of quantum synchronization has not yet been established, and different measures may characterize different aspects of coordinated quantum dynamics.

Superconducting quantum circuits provide a highly controllable platform for studying collective quantum dynamics. In circuit quantum electrodynamics, superconducting qubits can be strongly coupled to quantized microwave fields confined in on-chip resonators, allowing coherent qubit-cavity interactions to be engineered and measured with high precision~\cite{Blais2004,Majer2007,Devoret2013,Blais2021}. A common cavity mode can mediate effective interactions between qubits that do not directly couple to one another. This makes cavity-coupled superconducting qubits a natural system for exploring synchronization phenomena in the quantum regime.

Synchronization is closely related to, but distinct from, entanglement and other quantum correlations. Entanglement measures quantify nonlocal correlations among subsystems, whereas synchronization concerns whether local dynamical variables or reduced quantum states evolve in a coordinated manner. Previous studies have shown that synchronization and entanglement may coexist or influence each other, but they are not equivalent phenomena~\cite{Roulet2018,Eneriz2019,Chepelianskii2024}. Genuinely quantum effects, such as energy quantization and interference, can reshape synchronization behavior through mechanisms such as synchronization blockade and interference blockade~\cite{Loerch2017,Kehrer2024}. In open quantum systems, dissipation, environmental memory, and non-Markovian effects can further modify the emergence and stability of synchronization~\cite{Karpat2021,Gaidi2026}. Recent studies have also extended quantum synchronization to qubit-resonator and superconducting-circuit settings, including dissipative qubits coupled to a driven resonator~\cite{Chepelianskii2024}, qubit synchronization mediated by the dynamical Casimir effect~\cite{Mitarai2024}, noise-induced synchronization in a superconducting transmon-qubit chain~\cite{Tao2025}, and nonreciprocal quantum synchronization~\cite{Lai2025}. These developments demonstrate that synchronization in superconducting quantum systems can be shaped by dissipation, noise, coherence, and engineered coupling structures. Here, we focus on a complementary situation in which recurrent synchronization emerges from coherent cavity-mediated exchange in a three-qubit system.

In this work, we investigate recurrent pairwise synchronization in a system of three superconducting qubits coupled to a single-mode microwave cavity. The qubits do not interact directly with one another; instead, their dynamics are mediated by the common cavity field. Although similar cavity-coupled multiqubit architectures have been used to study entanglement generation and redistribution~\cite{Huan2022}, the present work shifts the focus from nonlocal correlations to the synchronization of local reduced states. The dressed-state formulation is used here as a standard dynamical framework for obtaining the coherent time evolution, while the central quantities of interest are the trace distances between reduced single-qubit density matrices. The synchronization considered here is therefore defined operationally in terms of reduced-state similarity. A smaller trace distance indicates that two qubits have more similar local quantum states, while a vanishing trace distance corresponds to complete synchronization of the reduced states. This definition does not rely on a classical phase-locking variable and is suitable for qubit systems in which local quantum states, rather than classical trajectories, are the relevant dynamical objects.

The trace-distance signals exhibit recurrent oscillatory wave-packet structures, indicating that synchronization appears within finite time windows rather than as a stationary steady state. To quantify the duration of the low-trace-distance synchronization windows, we apply an envelope-threshold method. By scanning the middle-qubit coupling strength and the relative coupling ratio of the two outer qubits, we construct a parameter-dependent map of the representative synchronization-window duration. Our results show that cavity-mediated coherent interactions can generate controllable recurrent pairwise synchronization among superconducting qubits, providing a trace-distance-based characterization of collective dynamics in multiqubit circuit-QED systems.

\section{Model and Dressed-State Formulation}\label{sec:model}

\begin{figure}[htbp]
\centering
\includegraphics[width=\textwidth]{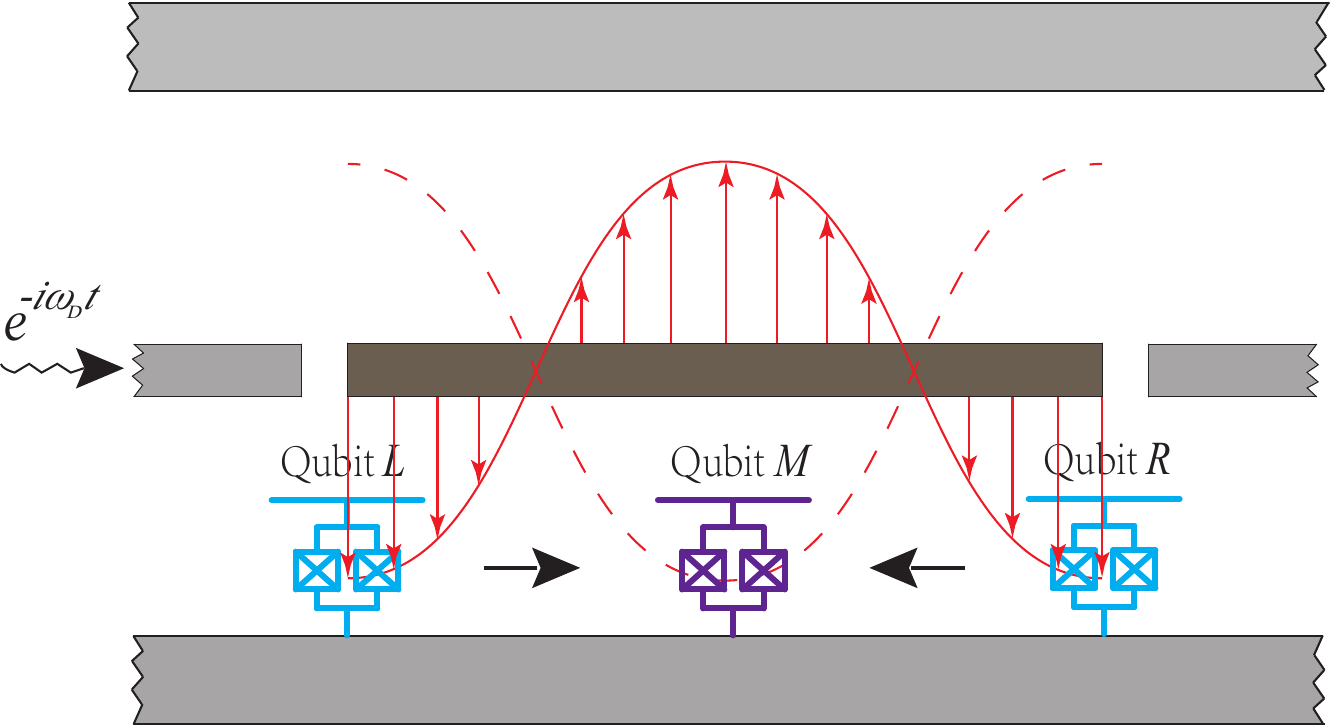}
\caption{Schematic illustration of three superconducting qubits coupled to a common single-mode microwave cavity. The middle qubit is fixed at a field antinode, while the two outer qubits are symmetrically positioned and their cavity couplings can be varied through their positions.}
\label{fig:system}
\end{figure}

The system consists of three superconducting qubits coupled to a single-mode microwave cavity, as illustrated in Fig.~\ref{fig:system}. The three qubits are labeled as $L$, $M$, and $R$, respectively. They do not interact directly with one another; instead, their dynamics are mediated by the common cavity field. The total Hamiltonian is
\begin{equation}
H=H_0+H_{\mathrm{int}}+H_d(t),
\label{eq:hamiltonian}
\end{equation}
where
\begin{align}
H_0 &= \omega_c a^\dagger a + \sum_{\nu=L,M,R}\Omega_\nu\sigma_{z,\nu}, \\
H_{\mathrm{int}} &= \sum_{\nu=L,M,R}\eta_\nu\left(a\sigma_{+,\nu}+a^\dagger\sigma_{-,\nu}\right), \\
H_d(t) &= i\varepsilon_D\left(a^\dagger e^{-i\omega_D t}-a e^{i\omega_D t}\right).
\end{align}
Here, $a$ and $a^\dagger$ are the annihilation and creation operators of the cavity mode with frequency $\omega_c$. The operators
\begin{equation}
\sigma_{z,\nu}=\frac{1}{2}\left(|e_\nu\rangle\langle e_\nu|-|g_\nu\rangle\langle g_\nu|\right)
\end{equation}
act on qubit $\nu$, whose transition frequency and qubit-cavity coupling strength are denoted by $\Omega_\nu$ and $\eta_\nu$, respectively. The operators $\sigma_{+,\nu}$ and $\sigma_{-,\nu}$ are the raising and lowering operators. The parameters $\varepsilon_D$ and $\omega_D$ are the amplitude and frequency of the coherent microwave drive. We denote the undriven Hamiltonian by $H_s=H_0+H_{\mathrm{int}}$.

In the numerical calculation, the cavity mode is truncated to the two lowest Fock states, $|0\rangle$ and $|1\rangle$. Together with the three two-level qubits, the total Hilbert space is therefore sixteen-dimensional. We denote the eight bare three-qubit states by $|\phi_m\rangle$, with $m=0,\ldots,7$, and write the full bare basis as
\begin{equation}
|\phi_m,n\rangle=|\phi_m\rangle\otimes|n\rangle,\qquad m=0,\ldots,7,\quad n=0,1.
\label{eq:bare-basis}
\end{equation}

To obtain the dressed-state representation, we diagonalize the undriven Hamiltonian,
\begin{equation}
H_s|u_k\rangle=E_k|u_k\rangle,\qquad k=0,\ldots,15.
\label{eq:dressed-eigenvalue}
\end{equation}
The dressed states are expanded in the bare basis as
\begin{equation}
|u_k\rangle=\sum_{m=0}^{7}\left[\alpha_{m,k}^{(0)}|\phi_m,0\rangle+\alpha_{m,k}^{(1)}|\phi_m,1\rangle\right],
\label{eq:dressed-state}
\end{equation}
where $\alpha_{m,k}^{(0)}$ and $\alpha_{m,k}^{(1)}$ are the components of the $k$th dressed state in the zero-photon and one-photon sectors. In the truncated basis, the annihilation operator is
\begin{equation}
a=\sum_{m=0}^{7}|\phi_m,0\rangle\langle\phi_m,1|.
\end{equation}
We define the cavity matrix element in the dressed-state basis as
\begin{equation}
\gamma_{ij}=\sum_{m=0}^{7}\alpha_{m,i}^{(0)*}\alpha_{m,j}^{(1)}.
\label{eq:gamma}
\end{equation}

The time-dependent state vector is expanded as
\begin{equation}
|\psi(t)\rangle=\sum_{i=0}^{15}c_i(t)|u_i\rangle.
\end{equation}
Substituting this expansion into the time-dependent Schr\"odinger equation gives
\begin{equation}
\dot{c}_i(t)=-iE_i c_i(t)+\varepsilon_D\sum_{j=0}^{15}\left[\gamma_{ji}^*e^{-i\omega_D t}-\gamma_{ij}e^{i\omega_D t}\right]c_j(t).
\label{eq:schrodinger-coefficients}
\end{equation}
These coupled equations are solved numerically to obtain the dressed-state amplitudes $c_i(t)$. The evolved state is then transformed back to the bare qubit-cavity basis,
\begin{equation}
|\psi(t)\rangle=\sum_{m=0}^{7}\left[\beta_m^{(0)}(t)|\phi_m,0\rangle+\beta_m^{(1)}(t)|\phi_m,1\rangle\right],
\end{equation}
with
\begin{equation}
\beta_m^{(n)}(t)=\sum_{k=0}^{15}c_k(t)\alpha_{m,k}^{(n)},\qquad n=0,1.
\end{equation}
The coefficients $\beta_m^{(n)}(t)$ are used to construct the full density matrix and the reduced density matrices of the individual qubits.

Unless otherwise stated, the cavity frequency is chosen as $\omega_c/2\pi=5.13~\mathrm{GHz}$ and the transition frequencies are $\Omega_L/2\pi=\Omega_R/2\pi=5.112~\mathrm{GHz}$ and $\Omega_M/2\pi=5.111~\mathrm{GHz}$. The cavity is driven by an external coherent microwave field with $\omega_D/2\pi=6.11~\mathrm{GHz}$ and $\varepsilon_D/2\pi=20~\mathrm{MHz}$. The drive is detuned from the bare cavity frequency and is kept fixed throughout the parameter scan. The initial state is the normalized coherent superposition
\begin{equation}
|\psi(0)\rangle=\sqrt{0.8}|e_L,g_M,g_R,0\rangle+\sqrt{0.1}|g_L,g_M,g_R,1\rangle+\sqrt{0.1}|e_L,g_M,g_R,1\rangle.
\label{eq:initial-state}
\end{equation}
We impose symmetric outer-qubit couplings, $\eta_L=\eta_R=\lambda\eta_M$, and scan $300\leq\eta_M/2\pi\leq500~\mathrm{MHz}$ and $0.05\leq\lambda\leq1.0$.

\section{Trace-Distance Synchronization Measure and Interval Extraction}\label{sec:measure}

To quantify synchronization between different qubits, we use the trace distance between their reduced single-qubit density matrices. After the time evolution is obtained, the full density matrix of the qubit-cavity system is constructed as $\rho(t)=|\psi(t)\rangle\langle\psi(t)|$. The reduced density matrices of the three qubits are obtained by tracing out the other degrees of freedom,
\begin{align}
\rho_L(t)&=\operatorname{Tr}_{MRC}[\rho(t)], &
\rho_M(t)&=\operatorname{Tr}_{LRC}[\rho(t)], &
\rho_R(t)&=\operatorname{Tr}_{LMC}[\rho(t)],
\end{align}
where $C$ denotes the cavity mode. The asynchronization between two qubits $\mu$ and $\nu$ is defined as
\begin{equation}
D_{\mu\nu}(t)=\frac{1}{2}\operatorname{Tr}\left|\rho_\mu(t)-\rho_\nu(t)\right|,\qquad \mu,\nu\in\{L,M,R\}.
\label{eq:trace-distance}
\end{equation}
The trace distance satisfies $0\leq D_{\mu\nu}(t)\leq1$. A smaller value indicates that the two qubits have more similar reduced quantum states. In particular, $D_{\mu\nu}(t)=0$ means that $\rho_\mu(t)=\rho_\nu(t)$, corresponding to complete synchronization in the reduced-state sense. Thus, $D_{\mu\nu}(t)$ is used as an asynchronization measure, and synchronization windows are intervals during which it remains close to zero.

The trace-distance signals generally exhibit recurrent oscillatory wave-packet structures. For each time trace $D_{\mu\nu}(t)$, we construct a smoothed envelope by applying a moving-average filter to the absolute trace distance,
\begin{equation}
E_{\mu\nu}(t)=\operatorname{Smooth}\left[|D_{\mu\nu}(t)|\right].
\label{eq:envelope}
\end{equation}
Since the amplitude varies with the system parameters, the threshold is chosen adaptively for each trace:
\begin{equation}
D_{\mathrm{th}}=P_{10}+\chi\left(P_{90}-P_{10}\right),
\label{eq:threshold}
\end{equation}
where $P_{10}$ and $P_{90}$ are the 10th and 90th percentiles of $E_{\mu\nu}(t)$, respectively, and $\chi$ is a dimensionless threshold parameter. In the present analysis, $\chi=0.05$ is used.

The high-trace-distance wave packets are identified by $E_{\mu\nu}(t)>D_{\mathrm{th}}$. The low-envelope regions between adjacent high-trace-distance wave packets correspond to low-trace-distance synchronization windows. If $t_r^{\mathrm{end}}$ is the end time of the $r$th wave packet and $t_{r+1}^{\mathrm{start}}$ is the start time of the subsequent wave packet, the duration of the intervening synchronization window is
\begin{equation}
\Delta t_r=t_{r+1}^{\mathrm{start}}-t_r^{\mathrm{end}}.
\label{eq:window-duration}
\end{equation}
When multiple synchronization windows occur within one trace, we select the longest one as the representative synchronization-window duration,
\begin{equation}
\Delta t=\max_r\left(\Delta t_r\right).
\label{eq:representative-duration}
\end{equation}
This choice characterizes the longest continuous interval over which two reduced qubit states remain close to each other while reducing the influence of small residual oscillations in the low-trace-distance regions. For each parameter set, $\Delta t$ is used to construct a parameter-dependent map as a function of the middle-qubit coupling strength and the relative coupling ratio of the outer qubits.

\section{Recurrent Pairwise Synchronization Dynamics}\label{sec:results}

\begin{figure}[htbp]
\centering
\includegraphics[width=0.82\textwidth]{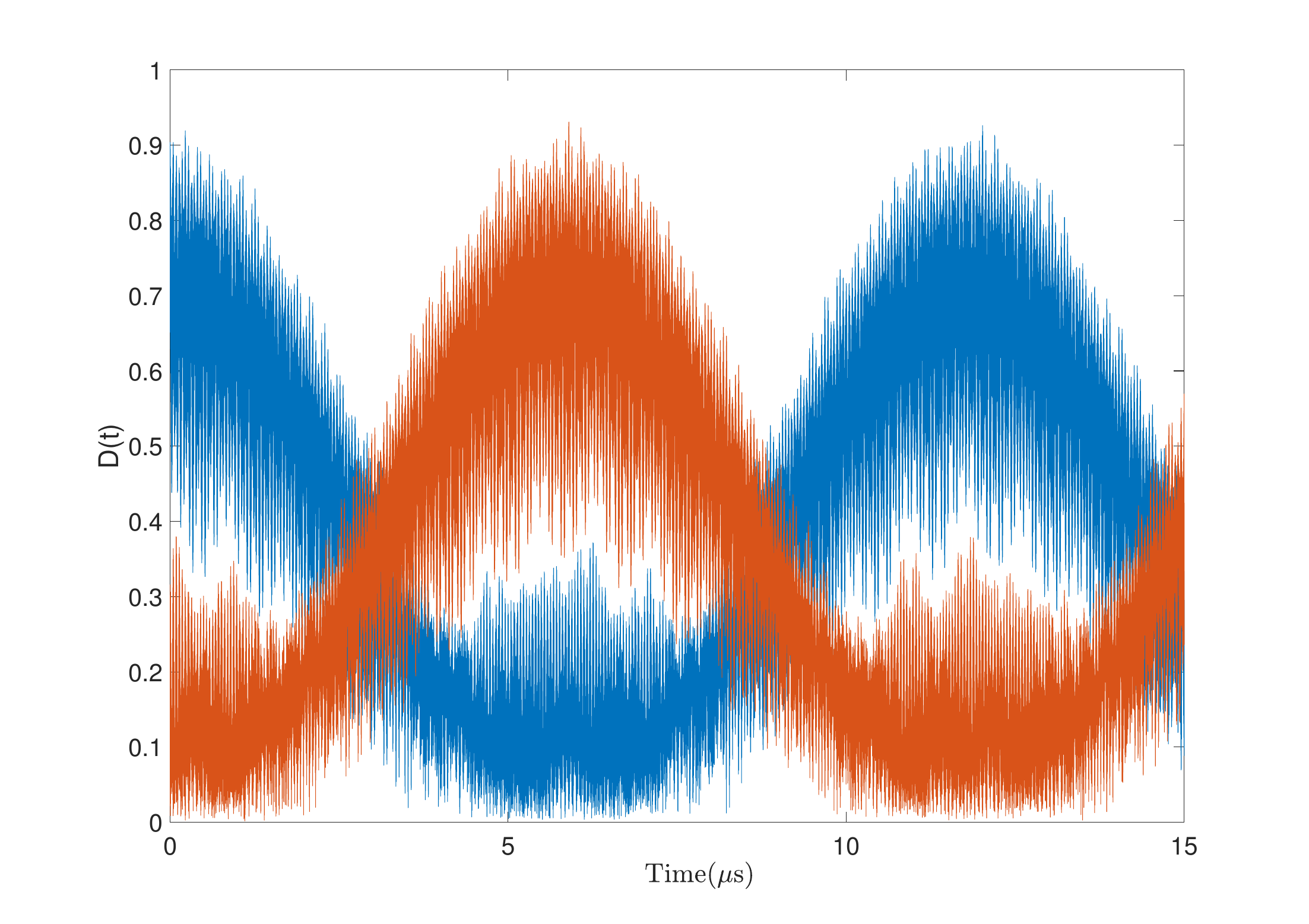}
\caption{Time evolution of $D_{LM}(t)$ and $D_{MR}(t)$ for $\eta_M/2\pi=300~\mathrm{MHz}$ and $\lambda=0.2$. The blue and red curves represent $D_{LM}(t)$ and $D_{MR}(t)$, respectively.}
\label{fig:trace-distance}
\end{figure}

We first examine a representative time evolution to illustrate recurrent pairwise synchronization dynamics. Figure~\ref{fig:trace-distance} shows the trace-distance signals $D_{LM}(t)$ and $D_{MR}(t)$ for $\eta_M/2\pi=300~\mathrm{MHz}$ and $\lambda=0.2$. Both signals exhibit pronounced oscillatory wave-packet structures, indicating that reduced-state similarity between different qubit pairs emerges repeatedly within finite time windows rather than as a stationary state. When $D_{LM}(t)$ is close to zero, the reduced states of the left and middle qubits become nearly identical, corresponding to an $L$-$M$ synchronization window. Similarly, low values of $D_{MR}(t)$ indicate synchronization between the middle and right qubits. The low-$D_{LM}$ and low-$D_{MR}$ regions occur in an alternating manner, rather than being minimized simultaneously. This behavior is consistent with a coherent cavity-mediated redistribution of local reduced-state similarity among the three qubits, with the middle qubit participating in alternating pairwise synchronization with the two outer qubits.

\begin{figure}[htbp]
\centering
\includegraphics[width=\textwidth]{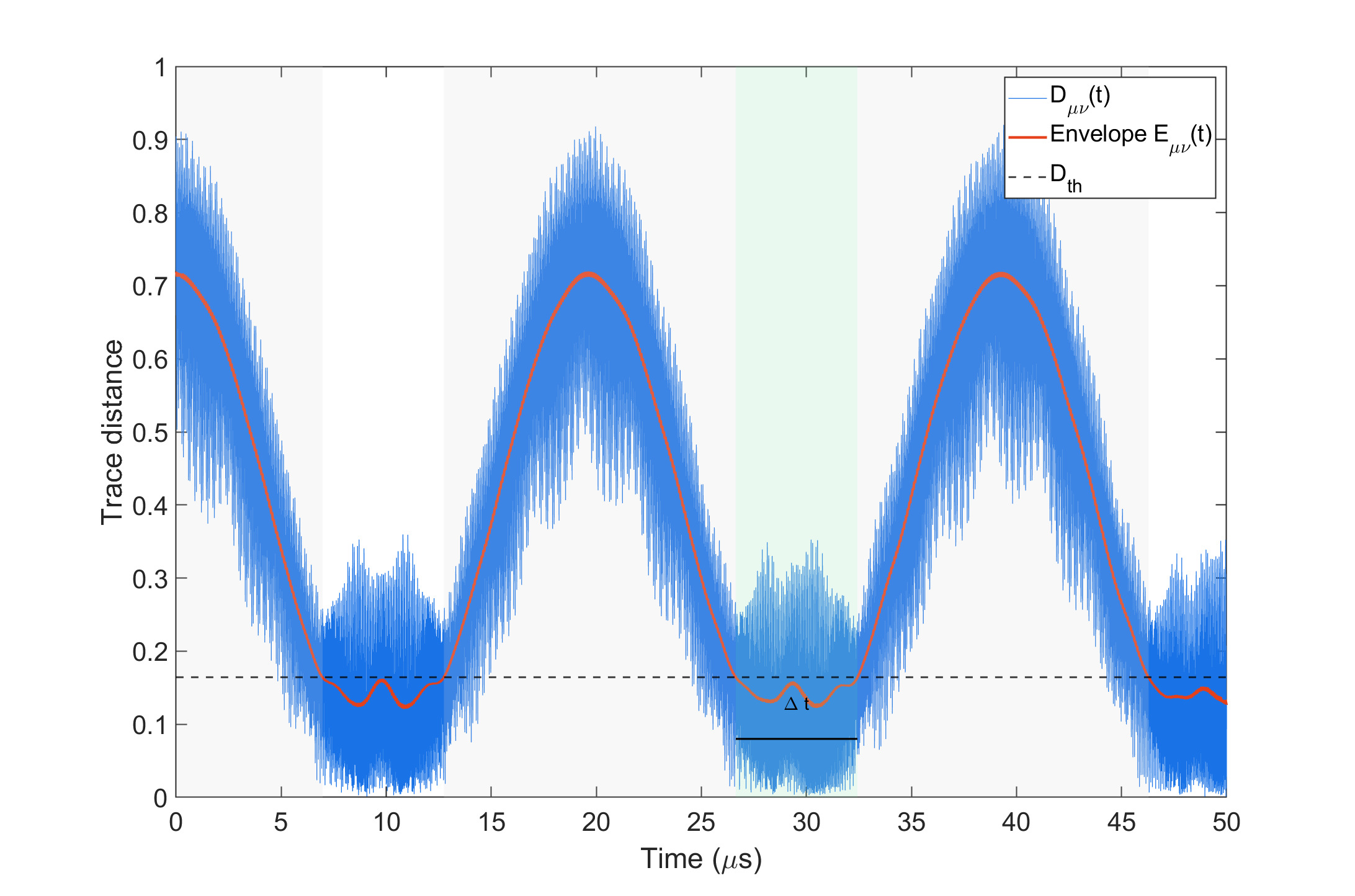}
\caption{Illustration of the envelope-threshold method for extracting the synchronization-window duration. The original trace-distance signal $D_{\mu\nu}(t)$ is shown in blue, and the smoothed envelope $E_{\mu\nu}(t)$ is shown in orange. The adaptive threshold $D_{\mathrm{th}}$ separates the high-trace-distance wave packets from the low-envelope regions. The highlighted interval $\Delta t$ is the duration of a low-trace-distance synchronization window.}
\label{fig:extraction}
\end{figure}

To quantify the time scale of the low-trace-distance regions, we apply the envelope-threshold procedure to a representative $D_{LM}(t)$ trace. Figure~\ref{fig:extraction} shows the extraction process for $\eta_M/2\pi=300~\mathrm{MHz}$ and $\lambda=0.15$. Although the raw trace contains rapid small-amplitude oscillations, the smoothed envelope $E_{LM}(t)$ reveals the slowly varying wave-packet structure. The adaptive threshold $D_{\mathrm{th}}$ separates the main high-trace-distance wave packets from the low-envelope regions. The shaded region marks a low-envelope synchronization window, and its duration $\Delta t$ is the separation between the end of one main wave packet and the beginning of the next. All adjacent intervals are identified from the envelope, and the largest one is selected as the representative duration for the corresponding parameter set.

\begin{figure}[htbp]
\centering
\includegraphics[width=\textwidth]{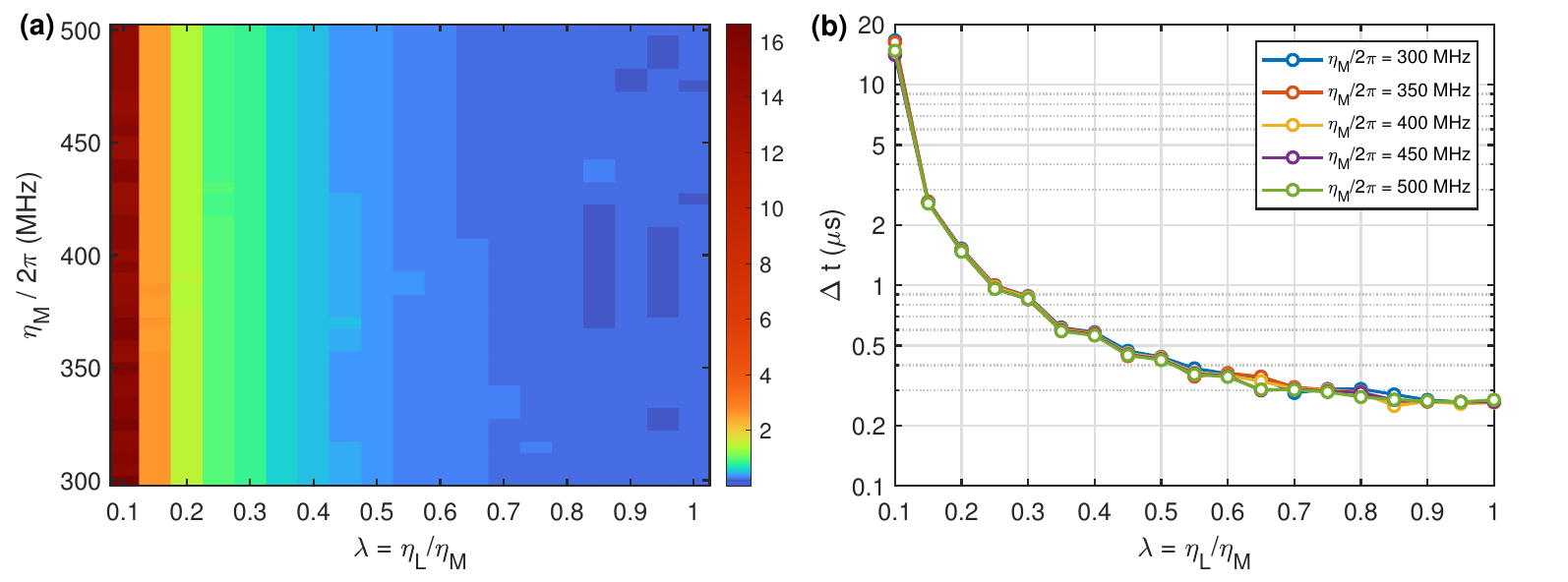}
\caption{Synchronization-window duration as a function of the coupling configuration. (a) Parameter-dependent map of the representative duration $\Delta t$ obtained from $D_{LM}(t)$ as a function of the relative coupling ratio $\lambda=\eta_L/\eta_M=\eta_R/\eta_M$ and the middle-qubit coupling strength $\eta_M/2\pi$. The color scale represents $\Delta t$ in microseconds. (b) Representative line cuts of $\Delta t$ as a function of $\lambda$ for $\eta_M/2\pi=300$, $350$, $400$, $450$, and $500~\mathrm{MHz}$.}
\label{fig:duration-map}
\end{figure}

After extracting the synchronization-window duration for all parameter sets, we construct the parameter-dependent map in Fig.~\ref{fig:duration-map}(a). The horizontal axis represents the relative coupling ratio $\lambda=\eta_L/\eta_M=\eta_R/\eta_M$, while the vertical axis denotes the middle-qubit coupling strength $\eta_M/2\pi$, varied from $300$ to $500~\mathrm{MHz}$. The map is constructed from the $D_{LM}(t)$ traces, and the color scale gives $\Delta t$ in microseconds. A pronounced decrease of $\Delta t$ is observed as $\lambda$ increases, indicating that larger relative coupling of the outer qubits to the cavity reduces the duration of low-trace-distance synchronization windows.

This trend is further illustrated by the representative line cuts in Fig.~\ref{fig:duration-map}(b). For different values of $\eta_M/2\pi$, the extracted duration is relatively large in the weak outer-coupling regime and decreases rapidly as $\lambda$ increases. At sufficiently large $\lambda$, the curves converge to a small duration and become weakly dependent on $\eta_M$. This behavior suggests that the synchronization-window duration is mainly controlled by the relative coupling ratio in the weak-coupling regime, whereas the dynamics become less sensitive to further changes in the coupling configuration in the stronger-coupling regime. Physically, increasing $\lambda$ enhances the participation of the outer qubits in cavity-mediated exchange and shortens the duration of intervals in which the associated reduced states remain close.

For the smallest coupling ratio, $\lambda=0.05$, fewer than two envelope-defined wave packets are captured for a substantial fraction of the scanned $\eta_M$ values within the $50~\mu\mathrm{s}$ observation window. These parameter sets are therefore excluded from the quantitative duration map and representative line cuts.

\clearpage
\section{Conclusion}\label{sec:conclusion}

We have studied recurrent pairwise synchronization in a coherently driven system of three superconducting qubits coupled to a common microwave cavity. By calculating the reduced density matrices of the individual qubits and evaluating their pairwise trace distances, we characterized synchronization operationally as the recurrent approach of two local qubit states. The resulting trace-distance dynamics exhibit oscillatory wave-packet structures, showing that synchronization occurs within finite time windows rather than as a stationary state.

To quantify this behavior, we introduced an envelope-threshold procedure for extracting the representative duration $\Delta t$ of a low-trace-distance synchronization window. The parameter-dependent results show that this duration decreases rapidly as the relative outer-qubit coupling ratio $\lambda=\eta_L/\eta_M=\eta_R/\eta_M$ increases and approaches a weakly coupling-dependent value at larger $\lambda$. Thus, stronger outer-qubit participation in the cavity-mediated exchange process reduces the duration of the low-trace-distance synchronization windows.

Our results identify trace distance as a direct reduced-state-based tool for characterizing coherent synchronization dynamics in multiqubit circuit-QED systems. They complement entanglement-based studies of the same cavity-mediated architecture by revealing recurrent synchronization between different qubit pairs. Extending the present analysis to include dissipation, dephasing, and experimentally realistic control imperfections will be important for assessing the observability and robustness of these synchronization windows in superconducting-circuit experiments.

\section*{Acknowledgements}
This work was supported by the Natural Science Foundation of Inner Mongolia Autonomous Region (Grant No.~2025QN01026) and the Research Program of Science and Technology at Universities of Inner Mongolia Autonomous Region (Grant No.~NJZY23031).

\section*{Data availability}
The numerical data generated and analyzed during the current study are available from the corresponding author upon reasonable request.

\section*{Code availability}
The numerical codes used in this study are available from the corresponding author upon reasonable request.

\section*{Conflict of interest}
The author declares no competing interests.

\bibliographystyle{unsrt}
\bibliography{references}

\end{document}